\documentclass[11pt]{article}
\usepackage{amsmath,amssymb,epsfig,cite}
\usepackage{dsfont}
\usepackage[all]{xy}
\usepackage{graphicx}
\usepackage{fancyhdr}

-1.1truecm \advance\voffset by -1.0truecm

\newcommand{\be}{\begin{equation}}
\newcommand{\ee}{\end{equation}}
\newcommand{\bea}{\begin{eqnarray}}
\newcommand{\eea}{\end{eqnarray}}

\newcommand{\pp}{\hat{\mathcal{P}}_+^\uparrow}
\newcommand{\at}{\mathcal{A}_\theta(\mathbb{R}^d)}

\numberwithin{equation}{section}

\begin{document}

\title{Poincar\'e Quasi-Hopf Symmetry and Non-Associative Spacetime
Algebra from Twisted Gauge Theories}

\bigskip

 \author{ A. P. Balachandran$^{a,b}$\footnote{C\'atedra de Excelencia}\ \footnote{bal@phy.syr.edu}
  and B. A. Qureshi$^{c,d}$\footnote{bqureshi@stp.dias.ie}\\
  \\
 $^a$\begin{small}{\it Department of Physics, Syracuse University,
Syracuse NY, 13244-1130, USA.} \end{small}\\
$^b$\begin{small}{\it Departamento de Matem\'{a}ticas, Universidad
Carlos III de Madrid, 28911 Legan\'{e}s, Madrid, Spain.}
\end{small}\\
$^c$ \begin{small} {\it School of Theoretical Physics, Dublin
Institute for Advanced Studies, Dublin 4, Ireland.}
\end{small}\\
$^d$ \begin{small} {\it Perimeter Institute for Theoretical
Physics, Waterloo, Ontario, N2L2Y5, Canada}
\end{small}}

\date{\empty}

\maketitle

\thispagestyle{fancyplain}

\fancyhf{}

\rhead{\fancyplain{}{SU-4252-886\\
DIAS-STP-09-12\\
 PI-other-148}} \cfoot{\fancyplain{}{\thepage}}

\begin{abstract}
In previous work, starting from the Moyal plane, we formulated
interacting theories of matter and gauge fields with only the
former fields twisted. In this approach, gauge theories, including
the standard model, can be formulated without new gauge degrees of
freedom. We show their underlying symmetry algebra to be
Poincar\'e quasi-Hopf . The associated spacetime algebra is hence
non-associative.
\end{abstract}

\section{Introduction}

When spacetime is noncommutative, it is often the case that
diffeomorphisms do not act as a {\it group} of automorphisms of
this algebra. Instead it can be the case that symmetries act on
the spacetime algebra as a {\it Hopf} or a {\it quasi-Hopf}
algebra \cite{Dimitrijevic:2004rf, Aschieri:2005yw,
Chaichian:2004za}. A prominent example is provided by the
Groenewald-Moyal (GM) plane $\mathcal{A}_\theta(\mathbb{R}^d)$ and
the Poincar\'e symmetry. The algebra
$\mathcal{A}_\theta(\mathbb{R}^d)$ is the algebra of functions on
$\mathbb{R}^d$ with the ``$\ast$'' product

\begin{align}
f_1\ast f_2=& f_1
e^{\frac{i}{2}\overleftarrow{\partial_\mu}\theta^{\mu\nu}\overrightarrow{\partial_\nu}}f_2,\\
f_i\ \in\  \mathcal{A}_\theta(\mathbb{R}^d),&\ \
\theta^{\mu\nu}=-\theta^{\nu\mu}=\textrm{ constant}.\nonumber
\end{align}
Let $\hat{\mathcal{P}}_+^\uparrow$ be the standard universal cover
of the (let us say) the connected Poincar\'e group. Then
$\hat{\mathcal{P}}_+^\uparrow$ does not act as a standard group of
automorphisms on $\at$ since $\theta^{\mu\nu}$ are constants. There
is however a Hopf algebra $(\mathbb{C}\pp, \Delta_\theta)$ where
$\mathbb{C}\pp$ is the group algebra of $\pp$ and $\Delta_\theta$ is
a deformed coproduct:
\begin{align}
\Delta_\theta(g)&=F_\theta^{-1}g\otimes g F_\theta,\ g\in \pp\quad,\\
F_\theta&= e^{\frac{i}{2}\partial_\mu\theta^{\mu\nu}\otimes
\partial_\nu}=\textrm{Drinfel'd's twist factor}\quad .
\end{align}
(We do not include the counit $\epsilon$ and the antipode $S$ in the
notation for simplicity).

The twist of coproduct implies the twist of statistics as well. Its
effects can be accounted for by ``dressing"
\cite{Zamolodchikov:1978xm,Grosse:1977ua,
Faddeev:1980zy} the quantum field $\phi_0$ of matter for
$\theta^{\mu\nu}$=0:
\begin{align}
\phi_\theta &= \textrm{quantum field of matter for noncommutativity
parameter} \ \theta^{\mu\nu}\nonumber\\
&=\phi_0
e^{\frac{1}{2}\overleftarrow{\partial_\mu}\theta^{\mu\nu}P_\nu},\
P_\nu =\textrm{Total momentum of {\it all} fields.}\nonumber
\end{align}
(Although here we focus on just one field, this formula is valid
in an interacting field theory with many fields. Then $P_\nu$
refers to the full four-momentum of the interacting field theory.)

Hereafter $A\wedge B$ will denote $A_\mu\theta^{\mu\nu}\otimes
B_\nu$.

A remarkable feature of the dressing transformation is its
self-reproducing  property:
\begin{equation}
\phi_\theta\ast\chi_\theta=(\phi_0\chi_0)e^{\frac{1}{2}\overleftarrow{\partial}\wedge
P}.
\end{equation}
In particular, for the interaction Hamiltonian density, it implies
\cite{Balachandran:2005pn, Balachandran:2006pi} that,
\begin{equation}
\mathcal{H}_I^\theta=\mathcal{H}_I^0
e^{\frac{1}{2}\overleftarrow{\partial}\wedge P}
\end{equation}
and that the interaction representation $S$-operator is
independent of $\theta^{\mu\nu},\ S_\theta=S_0$. But  scattering
amplitudes show time delays which depend on $\theta^{\mu\nu}$
\cite{Balachandran:2005pn,Buchholz:2008rd}.

The above approach has no major physical problems in the absence of
gauge fields. When gauge fields are introduced, new issues arise. In
the covariant derivative , $D_\mu=\partial_\mu+A_\mu,$ at first
sight, it seems natural to regard $A_\mu$ as $\underline{G}$-valued
functions on $\at$ where $\underline{G}$ is the Lie algebra of the
compact simple group $G$ underlying the gauge theory. Unfortunately,
as is well-known, this point of view cannot be sustained , since
$[D_\mu, D_\nu]$ is valued in the enveloping algebra
$\mathcal{U}(\underline{G})$ of $\underline{G}$. If we work in an
$N$-dimensional irreducible representation of $\underline{G}$,
$[D_\mu, D_\nu]$ is generally valued in $\underline{U(N)}$. One {\it may} thus be obliged
to introduce new gauge fields \cite{Aschieri:2006ye} causing
problems in formulating for example the standard model on $\at$.

We remark however that new gauge degrees of freedom may not be necessary. Vassilevich \cite{Vassi} has found new gauge invariant expressions which vanish as $\theta^{\mu\nu}\to 0$ and which can be added to the action. With their inclusion, it may be possible to avoid new gauge degrees of freedom.

In past work \cite{Balachandran:2007kv, Balachandran:2007yf}, we
developed an alternative formulation. There the gauge fields
$A_\mu$ are $\underline{G}$-valued functions on the commutative
algebra $\mathcal{A}_0(\mathbb{R}^d)$. The fields $A_\mu$
are thus not twisted: $A_\mu^\theta=A_\mu^0$. Matter fields are still based
on $\at$ and are given by $\phi_\theta$ where $P_\nu$ is now the
total momentum including that of gauge fields.

Such a formulation is possible since $\at$ is an
$\mathcal{A}_0(\mathbb{R}^d)$-module. It has specific consequences
such as the appearance of new types of diagrams, UV-IR mixing of a
new sort and CPT violation
\cite{Balachandran:2007yf,Jo,Invariance,Anosh}.

Thus gauge fields are based on the commutative algebra of functions
$\mathcal{A}_0(\mathbb{R}^d)$. Hence Poincar\'e transformations act
on gauge fields with the untwisted coproduct $\Delta_0$. The
corresponding P$\widehat{{\rm oincar}}$\'e Hopf algebra is $(\mathds{C}\pp,
\Delta_0)$ whereas it is $(\mathds{C}\pp, \Delta_\theta)$ for
matter fields. (The hat on P$\widehat{{\rm oincar}}$\'e is to show that we deal with its covering
group.)

As gauge and matter fields interact, the existence of two
different P$\widehat{{\rm oincar}}$\'e Hopf algebras raises
consistency questions regarding our treatment of Poincar\'e
symmetry. In this paper, we formulate a {\it single}
P$\widehat{{\rm oincar}}$\'e quasi-Hopf symmetry acting on both
matter and gauge fields
\cite{Mack:1991sr,Mack:1991tg,Mack:1992ez,Majid}. The coproduct on
this symmetry algebra is not coassociative. As a result, the
product on the spacetime algebra is not associative. The
statistics group too is changed: it is neither the permutation nor
the braid group.

Quasi-Hopf algebras were formulated by Drinfel'd. They were later
studied by Mack, Schomerus
\cite{Mack:1991sr,Mack:1991tg,Mack:1992ez}, Majid \cite{Majid} and
others. But perhaps it is here that they appear for the first time
in  the context of relativistic quantum field theories.

In this note, we describe the
preceding new results indicating all the necessary steps. But there
are several aspects not elaborated here such as the properties of
the $\mathcal{R}$-matrix and the construction of ``covariant products of quantum
fields" \cite{Mack:1991sr,Mack:1991tg,Mack:1992ez}. Elsewhere we will give a full treatment basing our
considerations on the work of Mack and Schomerus \cite{Mack:1991sr,Mack:1991tg,Mack:1992ez}.
But, for now, in the interests of simplicity, we highlight just the
main points.

This paper has been written with the Lehmann-Symanzik-Zimmermann
(LSZ) formalism of quantum field theories (qft's) on
$\mathcal{A}_{\theta}(\mathbb{R}^d)$ in \cite{Balachandran:2009gx}
in mind. It works with interacting fields and total
energy-momentum operators $P_{\mu}$ which include interactions.
But it is easily adapted to the perturbative approach of
\cite{Balachandran:2007kv} by replacing $P_{\mu}$ by their
free-field counterparts.

\section{The Drinfel'd Twists and Quasi-Hopf Algebras}\label{sec2}

Drinfel'd gives a general procedure to obtain new Hopf algebras
starting from a given Hopf algebra using twists. The construction
of the coproduct $\Delta_\theta$ is an example of this general
theory of twisting.

This section follows the treatment of Drinfel'd's work as given in
\cite{Majid}. We always assume that a quasitriangular structure
(the $\mathcal{R}-$matrix) exists. Here we only give the
definitions and properties which are essential to follow the later
sections for completeness. For details, see \cite{Majid}.

Consider a Hopf algebra $H$ with a coproduct $\Delta$, which acts
in another algebra $\mathcal{A}$ with  multiplication map $m_0$.
Now consider an invertible element $F\in H\otimes H$ ( the twist
element) which is a counital 2-cocycle ( a condition which we will
describe shortly). Then one can define a new Hopf algebra with the
same algebra structure as $H$, but with the new coproduct

\begin{equation}
\Delta_F=F^{-1}\Delta F,
\end{equation}
and this algebra acts in a new carrier algebra $\mathcal{A}_F$
where the multiplication rule is now given by
\begin{equation}
m_F=m_0F.
\end{equation}

The new coproduct is generally not cocommutative (even if the
original untwisted coproduct is) i.e. if we flip the entries in
the tensor product which appears in $\Delta_F(\cdot)$, we do not
get back the original coproduct:
\begin{equation}
\Delta'_F\equiv s\Delta_F\neq\Delta_F
\end{equation}
where $s$ is the transposition map which  flips the entries in the
tensor product in the coproduct. Hence the usual
symmetrization/antisymmetrization in the tensor products of the
carrier algebra ( that is, the statistics) is not compatible with
the coproduct. Rather, in any theory with multiparticle states,
the statistics is governed by the $\mathcal{R}-$matrix associated
with the coproduct.

The $\mathcal{R-}$matrix has the property

\begin{equation}
\mathcal{R}\Delta=\Delta'\mathcal{R}.\label{rmatrix}
\end{equation}
Therefore the correct statistics operator $\tau$ on
$\mathcal{A}\otimes \mathcal{A}$ which is compatible with a
general coproduct is given by
\begin{equation}
\tau=\sigma\circ(\rho\otimes\rho)(\mathcal{R}).
\end{equation}
Here $\sigma$ is the flip operator on the tensor product $V\otimes
V$ of representation carrier space $V$ and $\rho$ is a
representation by which $H$ acts in $V$.  The diagonalization with
respect to $\tau$ gives states which are superselected.

It is easy to see that the $\mathcal{R}-$matrix for the coproduct
obtained by twisting procedure from a trivial coproduct is given
by
\begin{equation}
\mathcal{R}=F_{21}^{-1}F,\label{rtwist}
\end{equation}
where
\begin{equation}
F_{21}^{-1}=s F^{-1}.
\end{equation}
where $s$ again flips the entries in the tensor product of
$F^{-1}$. So $\tau$ can be written as
\begin{equation}
\tau=\sigma\circ(\rho\otimes\rho)( F_{21}^{-1} F).
\end{equation}
We will often omit the representation symbol $\rho$ when it is
clear from the context.

 Thus we see that the twisting procedure
works at three levels. It not only twists the coproduct of the
symmetry group and the product in the spacetime algebra, but it
also changes the usual bosonic/fermionic statistics to twisted
bosonic/fermionic statistics.

\subsection{Coassociativity and Quasi-Hopf Agebras}

The coassociativity of coproduct is defined by
\begin{equation}
(id\otimes \Delta)\Delta=(\Delta\otimes id)\Delta.
\end{equation}
By duality, this represents the associativity of the carrier
algebra \cite{Balachandran:2009st}.

Drinfel'd has defined more general algebraic structures where the
above condition fails to hold, called quasi-Hopf algebras.
However, this failure is controlled by an intertwiner $\phi \in
H\otimes H\otimes H$ ( fulfilling certain properties which we will
not discuss) such that
\begin{equation}
(id\otimes \Delta)\Delta(h)=\phi\big((\Delta\otimes
id)\Delta(h)\big)\phi^{-1}\label{quasi}
\end{equation}
for all $h\in H$. The definitions for antipode. counit and
quasi-triangular structure are also appropriately modified. But we
will not discuss those here as well. It is (\ref{quasi}), that is
the central element leading to the definition of quasi-Hopf
algebras.

These quasi-Hopf algebras can actually be obtained by twisting
with a twist element $F$ which is required to be counital i.e.,
\begin{equation}
(id\otimes \epsilon)F=(\epsilon\otimes
id)F=\mathds{1}\label{counital}
\end{equation}
where $\epsilon$ is the counit. It is the 2-cocycle condition on
the twist element $F$,
\begin{equation}
(F\otimes \mathds{1})\cdot(\Delta\otimes id)(F)=(\mathds{1}\otimes
F)\cdot (id\otimes \Delta)(F),\label{cocycle}
\end{equation}
which ensures the coassociativity of the twisted coproduct. If the
twist element $F$ does not fulfill this condition, the resulting
Hopf algebra is only a quasi-Hopf algebra. Notice that $F$ only
needs to obey (\ref{counital}) to qualify as a twist for a
resulting quasi-Hopf algebra.

It is important to note that even in a quasi-Hopf algebra, the
$\mathcal{R}-$ matrix still obeys (\ref{rmatrix}) and is still
obtained via (\ref{rtwist}) from the twist operator $F$. Hence the
twisted statistics for a twisted quasi-Hopf algebra is again given
by
\begin{equation}
\tau=\sigma F^{-1}_{21}F .
\end{equation}
Where we omitted the symbol $\circ$ after $\sigma$.

In general, a quasi-Hopf algebra is a complicated object. However,
if it is obtained from a twist $F$, it is easy to use it as all
the structures of quasi-Hopf algebras follow from this twist.

\section{The Twisted Fields}

Twisted fields such as $\phi_\theta$ contain all the information
on statistics, and hence the coproduct on the symmetry algebra
$\mathds{C}\pp$ and the product on spacetime algebra. This is
fully explained in \cite{Balachandran:2007yf,Balachandran:2007vx}.
Therefore we first focus on a uniform construction of the twisted
fields. For this purpose, we have to enlarge $\mathds{C}\pp$ by
introducing a central element $u$. We call the extended algebra as
$\overline{\mathds{C}\pp}$.

The central element $u$ is effectively a grading operator for the
quantum fields. It behaves like a pure group element under the
coproduct $\overline{\Delta_\theta}$, counit $\overline{\epsilon}$
and antipode $\overline{S}$ of the extended algebra. Thus
\begin{align}\label{Mario1}
\overline{\Delta_\theta}(u)&= u\otimes u,\\
\overline{\epsilon}(c)&=\mathds{1},\\
\overline{S}(u)&=u^{-1}.
\end{align}

The $\ast$-operator on $\mathds{C}\pp$ is extended to
$\overline{\mathds{C}\pp}$ by setting
\begin{equation}
u^{\ast}u=uu^{\ast}=\mathds{1}.
\end{equation}
It is thus a unitary element.

Let $\chi_0^g$ and $\chi_0^m$ generically denote a basic untwisted
gauge and matter field. The element $u$ acts on the fields by
conjugation as usual. This representation of $u$ on fields is
denoted by $Ad$. Thus
\begin{equation}
Ad\,u\ \chi_0^{g,m}\ :=\ u\chi_0^{g,m}u^{-1}.
\end{equation}

We set
\begin{equation}
Ad\,u\ \chi_0^g=+\chi_0^g\ \ ,\ \  Ad\,u\
\chi_0^m=-\chi_0^m.\label{set}
\end{equation}
Thus $u$ is a grading operator with $\chi_0^g$ being even and
$\chi_0^m$ being odd.

We complete the definition of $u$ in quantum field theory by
setting $u=\mathds{1}$ on vacuum:
\begin{equation}
u|0\rangle=|0\rangle.
\end{equation}

It follows from (\ref{set})  that
\begin{equation}
\delta_{Ad\,u,-1}\equiv\frac{1}{2}[\mathds{1}-Ad\,u]
\end{equation}
acts as $0$ on $\chi_0^g$ and identity on $\chi_0^m$:
\begin{equation}
\delta_{Ad\,u,-1}\chi_0^g=0\ ,\
\delta_{Ad\,u,-1}\chi_0^m=\chi_0^m.\label{charges}
\end{equation}
It is thus a projector. We avoid the use of $P$ in denoting it as
$P$ stands for the momentum operator elsewhere.

We set as usual
\begin{align}
Ad\Delta_0(u)&\equiv (Ad\otimes Ad) (u\otimes u)\\
&=Ad\,u\otimes Ad\,u.
\end{align}

We now write the twisted field $\chi_\theta^{g,m}$, which can be
matter or gauge, as
\begin{equation}
\chi_\theta^{g,m}=\chi_0^{g,m}e^{\frac{1}{2}\overleftarrow{\partial}\wedge
P(\overleftarrow{\delta_{Ad\,u,-1}})}
\end{equation}
where the left arrow indicates action on $\chi_0^{g,m}$.

In view of (\ref{charges}),
\begin{equation}
\chi_\theta^g=\chi_0^g\ ,\ \chi_\theta^m=\chi_0^m
e^{\frac{1}{2}\overleftarrow{\partial}\wedge P}.
\end{equation}
These are exactly what we want.

The representation $Ad$ extends to $\chi_\theta^{g,m}$ in a
natural way:
\begin{equation}
Ad\,u\ \chi_\theta^{g,m}=u\chi_\theta^{g,m}u^{-1}.
\end{equation}

{\it Remark:}

The introduction of a new element to convert a symmetry algebra
into a Hopf algebra has occurred before . Thus the SUSY algebra is
not Hopf. Now let $N_F$ be the fermion number and consider
$(-1)^{N_F}$. It is the grading operator, commuting with even and
anticommuting with odd SUSY generators. Mack and Schomerus
\cite{Mack:1991sr} extend SUSY to $\overline{\textrm{SUSY}}$ by
including this element and show that $\overline{\textrm{SUSY}}$,
unlike SUSY, is Hopf.

\section{The Coproduct $\overline{\Delta}_\theta$ on
$\overline{\mathds{C}\pp}$}

In the previous section, we did not specify the twisted coproduct
 $\overline{\Delta_\theta}$ on
$\mathds{C}\pp\subset\overline{\mathds{C}\pp}$. We take up that
task here.

We know that the coproduct on the gauge sector is just the usual
coproduct without any twist,
\begin{equation}
\overline{\Delta_\theta}\mid_{\textrm{Gauge
fields}}=\Delta_0\mid_{\textrm{Gauge fields}},
\end{equation}
where
\begin{equation}
\Delta_0(g)=g\otimes g
\end{equation}
for a Poincar\'e group element $g\in \overline{\mathds{C}\pp}$.
For the matter sector, the coproduct is given by,
\begin{align}
\overline{\Delta_\theta}\mid_{\textrm{Matter
fields}}&=\Delta_\theta=F_\theta^{-1}\Delta_0F_\theta,\\
F_\theta &=e^{-\frac{i}{2}P_\mu\theta^{\mu\nu}\otimes P_\nu}.
\end{align}

We want to write a coproduct which reduces to the corresponding
coproducts on each sector using a single twist operator:
\begin{equation}
\overline{\Delta_\theta}=\overline{\mathfrak{F}_{\theta}}^{-1}
\Delta_0 \overline{\mathfrak{F}_\theta}.
\end{equation}
In this way we will be defining a new Hopf symmetry structure on
the full theory. The twist operator
$\overline{\mathfrak{F}_{\theta}}$ which does this job is given by
\begin{equation}
\overline{\mathfrak{F}_{\theta}}=e^{-\frac{i}{2}P_\mu\theta^{\mu\nu}\otimes
P_\nu (\delta_{Ad\,u,-1}\otimes\mathds{1})}.\label{twistfact}
\end{equation}
It reduces to corresponding twist factors in the respective
sectors.

We can do a check that this is indeed the twist factor for our
coproduct. We know that for a field $\phi$ to carry a
representation of any coproduct $\Delta$, it must fulfill
\begin{equation}
U(g)\phi=\phi( U\otimes \overleftarrow{\rho})(id\otimes
S)\Delta(g),\label{coprodrep}
\end{equation}
where $S$ in the antipode (inverse for pure group elements),
$U(g)$ is the operator representative of $g$ on the Hilbert space
and $\rho$ is the  representation of the group on the field
$\phi$:
\begin{equation}
(\rho(g)\phi)(x)=\phi(g^{-1}x),\ \ \ g\in\pp.
\end{equation}
The argument for $\rho$ comes from the second factors in
$(id\otimes S)\Delta(g)$. They act as usual from left to right in
$\phi$. For the untwisted coproduct $\Delta_0=g\otimes g$,
(\ref{coprodrep}) produces the standard result,
\begin{equation}
U(g)\phi(x)U(g)^\dagger =\phi(gx)
\end{equation}

In previous papers \cite{Balachandran:2007yf,Balachandran:2007vx},
we have shown that the usual expressions for operators $U(g)$ in
terms of untwisted oscillators fulfill equation (\ref{coprodrep})
with twisted coproduct when acting on the twisted fields. Thus one
knows that the algebraic structure in $\overline{\mathds{C}\pp}$
is not changed by changing the coproduct. Also, since the
operators act on the same Hilbert space as before, it is expected
that the operators $U(g)$ do not change when written in terms of
untwisted oscillators because otherwise they will not satisfy the
$\overline{\mathds{C}\pp}$ algebra. But the remarkable fact is
that when they act on twisted fields, they reproduce the twisted
coproduct. In other words, the transformations of the twisted
fields with correct coproduct can be obtained by simply
transforming the untwisted field and $P$'s in the standard manner.

Let us show this for spinless fields and $g$ a Lorentz
transformation. The results for more general fields follow easily.
Consider the twisted field $\chi_\theta^{g,m}$ given by
\begin{equation}
\chi_\theta^{g,m}\equiv\chi_0^{g,m}e^{\frac{1}{2}\overleftarrow{\partial}\wedge
P\big(\overleftarrow{\delta}_{Ad\,u,-1}\big)}.
\end{equation}
We can explicitly write the generators of $\pp$ in terms of in ( or
out) fields. Thus since $P_\mu$ is time-independent, at least
formally, we have, on letting $x_0\longrightarrow -\infty$,
\begin{equation}
\chi_\theta^{g,m,in}=\chi_0^{g,m,in}e^{\frac{1}{2}\overleftarrow{\partial}\wedge
P\big(\overleftarrow{\delta}_{Ad\,u,-1}\big)}.
\end{equation}
The Poincar\'e generators ( of the fully interacting theory) have
the expansions as in the free field case, but in terms of
$\chi_0^{in}$. Therefore we can calculate how
$\chi_\theta^{g,m,in}$ transforms and that is enough to find  the
coproduct on $\mathds{C}\pp$.

Now, acting by $U(g)$ on $\chi_\theta^{g,m,in}$ and transforming
the untwisted field $\chi_0^{g,m,in}$ and $P_\mu$ in the standard
way, we have, for $g$ a Lorentz transformation
\begin{align}
U(g)\chi_\theta^{g,m,in}(x)&=U(g)\chi_0^{g,m,in}(x)e^{\frac{1}{2}\overleftarrow{\partial}\wedge
P\big(\overleftarrow{\delta}_{Ad\,u,-1}\big)}\\
&=\chi_0^{g,m,in}(gx)U(g)e^{\frac{1}{2}\overleftarrow{\partial}\wedge
P\big(\overleftarrow{\delta}_{Ad\,u,-1}\big)}\\
&=\chi_0^{g,m,in}(gx)e^{\frac{1}{2}(\overleftarrow{g\partial})\wedge
P\big(\overleftarrow{\delta}_{Ad\,u,-1}\big)}e^{-\frac{1}{2}(\overleftarrow{g\partial})\wedge
P\big(\overleftarrow{\delta}_{Ad\,u,-1}\big)}U(g)e^{\frac{1}{2}\overleftarrow{\partial}\wedge
P\big(\overleftarrow{\delta}_{Ad\,u,-1}\big)}\\
&=\chi_\theta^{g,m,in}(gx)e^{-\frac{1}{2}(\overleftarrow{g\partial})\wedge
P\big(\overleftarrow{\delta}_{Ad\,u,-1}\big)}U(g)e^{\frac{1}{2}\overleftarrow{\partial}\wedge
P\big(\overleftarrow{\delta}_{Ad\,u,-1}\big)}\label{last2}
\end{align}
Now if we recall that on any field $\phi$, the  representation is
\begin{equation}
\rho(g^{-1})\phi(x)=\phi(gx),\ \ \
\rho(P_\mu)\phi(x)=i\partial_\mu\phi(x),
\end{equation}
then (\ref{last2}) is exactly same as (\ref{coprodrep}) with
$\Delta_\theta=\overline{\mathfrak{F}_{\theta}}^{-1}(g\otimes
g)\overline{\mathfrak{F}_{\theta}}$. ( Note that $S$ is an
anti-homomorphism).

\section{On Lack of Coassociativity of Coproduct
$\overline{\Delta_\theta}$}\label{lack}

The coproduct $\overline{\Delta_\theta}$ is not coassociative. We
can see this by evaluating $(id\otimes
\overline{\Delta_\theta})\overline{\Delta_\theta}(g)$ and
$(\overline{\Delta_\theta}\otimes id )\overline{\Delta_\theta}(g)$
on vectors $e_p\otimes e_q\otimes e_r \in
V_{\textrm{Gauge}}\otimes V_{\textrm{Matter}}\otimes
V_{\textrm{Matter}}$ where $V_{\textrm{Gauge}}$ and
$V_{\textrm{Matter}}$ denote vector spaces with $u=1$ and $u=-1$
and $e_k\ \ (k=p,q,r)$ denote plane wave vectors :
$e_k(x)=e^{ik\cdot x}$. Hence $P_\mu e_k=k_\mu e_k$.

Consider $\overline{\Delta_\theta}(g)$:
\begin{align}
\overline{\Delta_\theta(g)}=&e^{\frac{i}{2}P_\mu\theta^{\mu\nu}\otimes
P_\nu \big(\delta_{u,-1}\otimes\mathds{1}\big)}(g\otimes g)\nonumber\\
&e^{-\frac{i}{2}P_\mu\theta^{\mu\nu}\otimes P_\nu\big(\delta_{u,-1}\otimes\mathds{1}\big)}\\
=&(g\otimes g)e^{\frac{i}{2}(\Lambda(g)P)_\mu\theta^{\mu\nu}\otimes
(\Lambda(g)P)_\nu\big(\delta_{u,-1}\otimes\mathds{1}\big)}\nonumber\\
&e^{-\frac{i}{2}P_\mu\theta^{\mu\nu}\otimes P_\nu
\big(\delta_{u,-1}\otimes\mathds{1}\big)}
\end{align}
where $\Lambda:g\longrightarrow \Lambda(g)$ is the homomorphism from
$SL(2,C)$ to $\mathcal{L}_+^\uparrow$.

Now apply $id\otimes \overline{\Delta_{\theta}}$ on the above vectors and collect the
exponentials with no $\Lambda(g) P$. They come from the last term :
\begin{align}
(id\otimes \overline{\Delta_{\theta}})&e^{-\frac{i}{2}P_\mu\theta^{\mu\nu}\otimes
P_\nu \big(\delta_{u,-1}\otimes\mathds{1}\big)}\nonumber\\
=&exp\{-\frac{i}{2}P_\mu\otimes(\mathds{1}\otimes\theta^{\mu\nu}
P_\nu+\theta^{\mu\nu}P_\nu\otimes \mathds{1})\times\nonumber\\
&\times (\delta_{u,-1}\otimes\mathds{1}\otimes\mathds{1})\}\quad.
\end{align}
Applying this to $e_p\otimes e_q\otimes e_r \in
V_{\textrm{Gauge}}\otimes V_{\textrm{Matter}}\otimes
V_{\textrm{Matter}}$, where the $V$'s denote the vector spaces for gauge fields or matter as indicated by subscripts,
\begin{equation}
\textrm{left-hand side acting on } e_p\otimes e_q\otimes e_r=e_p\otimes e_q\otimes e_r\quad.
\end{equation}

Also
\begin{align}
(\overline{\Delta_\theta}\otimes id)&e^{-\frac{i}{2}P_\mu\theta^{\mu\nu}\otimes
P_\nu \big(\delta_{u,-1}\otimes\mathds{1}\big)}e_p\otimes e_q\otimes e_r\nonumber\\
=&exp\{-\frac{i}{2}(P_\mu\otimes\mathds{1}+\mathds{1}\otimes
P_\mu)\otimes \theta^{\mu\nu}
P_\nu\times\nonumber\\
&\times (\delta_{(u\otimes u,-1)}\otimes\mathds{1})\}e_p\otimes e_q\otimes e_r\\
=&e^{-\frac{i}{2}(p+q)_\mu\theta^{\mu\nu}r_\nu}e_p\otimes
e_q\otimes e_r
\end{align}
so that
\begin{equation}\label{Bab1}
(id\otimes\overline{\Delta_\theta})\overline{\Delta_\theta}\neq(\overline{\Delta_\theta}\otimes
id)\overline{\Delta_\theta}.
\end{equation}

\section{The Algebra of Functions}
Let us denote it by $\mathcal{B}_\theta(\mathbb{R}^N)$. It has two
components, with gradings $ u=+1$ and $-1$:
\begin{equation}
\mathcal{B}_\theta(\mathbb{R}^N)=\mathcal{B}_\theta^{+1}(\mathbb{R}^N)\oplus\mathcal{B}_\theta^{-1}(\mathbb{R}^N).
\end{equation}

The $*$-product on functions $\alpha,\beta \in
\mathcal{B}_\theta(\mathbb{R}^N)$ is
\begin{equation}
\alpha\ast\beta=m_0[\overline{F_\theta}\ \alpha\otimes\beta]
\end{equation}
where $m_0$ is the point-wise multiplication map and
\begin{equation}
\overline{F_\theta}=e^{\frac{i}{2}\partial_\mu\otimes
\theta^{\mu\nu}\partial_\nu\{\delta_{u,-1}\otimes\mathds{1}\}}\quad.
\end{equation}
as follows in the standard manner from the coproduct. It is
easy to check that this product is not coassociative. The calculation is similar to the one leading to (\ref{Bab1}) Thus
\begin{equation}
e_p\ast(e_q\ast e_r)\neq (e_p\ast e_q)\ast e_r
\end{equation}
\begin{equation}
e_p\in \mathcal{B}_\theta^{+1}(\mathbb{R}^N),\ \ e_{q,r}\in
\mathcal{B}_\theta^{-1}(\mathbb{R}^N).
\end{equation}

The loss of associativity also follows using general
considerations and the nonassociativity of the coproduct
\cite{Mack:1991 sr, Mack:1991 tg, Mack:1992ez}.

\section{The Quasi-Hopf Structure of $\overline{\pp}$}

We have a new coproduct on $\overline{\pp}$ which is obtained by
twisting with the twist element
$\overline{\mathfrak{F}_{\theta}}$. For this twist to generate a
Hopf algebra, it must satisfy (\ref{counital}), which it does,
owing to the fact that
\begin{equation}
\epsilon(P_\mu)=0.
\end{equation}

Now we also saw that the resulting coproduct
$\overline{\Delta_\theta}$ is not coassociative. It means that the
resultant Hopf algebra is only a quasi-Hopf algebra.

Indeed, for the twist to generate a Hopf algebra, it must satisfy
(\ref{cocycle}). But we can show that the twist element
$\overline{\mathfrak{F}_{\theta}}$ does not satisfy it. By simple
algebra one can calculate that
\begin{align}
(\overline{\mathfrak{F}_{\theta}}\otimes\mathds{1})(\Delta_0\otimes
id)\overline{\mathfrak{F}_{\theta}}=exp[-\frac{i}{2}\theta^{\mu\nu}\big(&(P_\mu\otimes
\mathds{1}\otimes P_\nu+\mathds{1}\otimes P_\mu\otimes
P_\nu)(\delta_{u\otimes u,-1}\otimes\mathds{1})\nonumber\\
&+(P_\mu\otimes P_\nu\otimes\mathds{1})(
\delta_{u,-1}\otimes\mathds{1}\otimes\mathds{1})\big)].\label{2}
\end{align}
On the other hand
\begin{align}
(\mathds{1}\otimes\overline{\mathfrak{F}_{\theta}})(id\otimes
\Delta_0)\overline{\mathfrak{F}_{\theta}}=exp[-\frac{i}{2}\theta^{\mu\nu}\big(&(P_\mu\otimes
P_\nu\otimes \mathds{1}+P_\mu\otimes\mathds{1}\otimes
P_\nu)(\delta_{u,-1}\otimes\mathds{1}\otimes\mathds{1})\nonumber\\
&+(\mathds{1}\otimes P_\mu\otimes P_\nu)(\mathds{1}\otimes
\delta_{u,-1}\otimes\mathds{1})\big)].\label{1}
\end{align}

It is clear that (\ref{2}) and (\ref{1}) are not equal. The terms
involving $\delta_{(u\otimes u),-1}$ in (\ref{2}) are absent in
(\ref{1}).  Hence
\begin{equation}
\mathds{1}\otimes\overline{\mathfrak{F}_{\theta}}(id\otimes
\Delta_0)\overline{\mathfrak{F}_{\theta}}\neq\overline{\mathfrak{F}_{\theta}}\otimes\mathds{1}(\Delta_0\otimes
id)\overline{\mathfrak{F}_{\theta}}.
\end{equation}
Thus this twist does not give an ordinary Hopf algebra. But we do
get a quasi-Hopf algebra. For as we explained in section
\ref{sec2} , all one needs is the property (\ref{counital}) to get
a quasi-Hopf algebra.

Actually if (\ref{cocycle}) were satisfied,
$\overline{\Delta_\theta}$ would have been coassociative. Its
failure proved in section \ref{lack} thus already shows that
(\ref{cocycle}) is not fulfilled.

\section{Final Remarks}
We have shown the existence of a quasi-Hopf symmetry structure in
a quantum gauge field theory where only matter fields feel the
noncommutativity of spacetime. The coproduct is not coassociative
and the $\ast$-product on the (two-sheeted) spacetime is not
associative in such a theory int the presence of both matter and
gauge fields.

\section{Acknowledgements}

We want to thank Alberto Ibort and the Universidad Carlos III de Madrid for their wonderful hospitality and support.

A.P.B. thanks T. R. Govindarajan and the Institute of Mathematical Sciences, Chennai for very friendly hospitality as well.

A.P.B. is most grateful to Mario Martone for help in preparing the manuscript.

The work of A.P.B.was supported in part by DOE under the grant
number DE-FG02-85ER40231 and by the Department of Science and
Technology, India.

The work of B.Q. was supported by IRCSET fellowship and by
Perimeter Institute.

\end{document}